\documentclass[aps,pra,preprint]{revtex4}
\usepackage{graphicx}
\begin{document}
\baselineskip=24pt
\title{An accurate exchange energy functional in excited-state density 
functional theory }
\author{Prasanjit Samal and Manoj K. Harbola}
\affiliation{Department of Physics, Indian Institute of Technology,
 Kanpur U.P. 208016, India}

\begin{abstract}
An exchange energy functional is proposed and tested for obtaining
a class of excited-state energies using density-functional formalism.
The functional is the excited-state counterpart of the local-density
approximation functional for the ground-state.  It takes care of the
state-dependence of the energy functional and leads to highly accurate
excitation energies.
\end{abstract}
\maketitle

\newpage
\section{Introduction}
Success of density functional theory (DFT) \cite{gross1,parr1} for the 
ground-state calculations had prompted search 
\cite{ziegler,Gunnar,vonB,LP1,rkp,theo,olivi1,nagy} 
for similar theories for the excited-states.  Over the past decade, time-dependent 
density-functional theory (TDDFT) \cite{gross2} has become a standard tool 
\cite{casida,peters} for obtaining transition energies and the associated oscillator 
strengths. However, despite its widespread use, the theory is not without limitations.
For example, calculating excitation energies for double excitation of electrons still 
remains \cite{burke} a challenge in the TDDFT approach. At the same time, the charm
of getting the excitation energy as the difference between two total energies
remains. This is because one can choose the excited-state at will, promoting
as many electrons as one wishes to a set of chosen orbitals, calculate the
total corresponding energy and find the excitation energy by subtracting the
ground-state energy. Thus research in the direction of performing a Kohn-Sham
like calculation for the excited-states continues.
 
A ground-state like DFT approach to obtain the total energy
of an excited-state has been developed by G\"{o}rling \cite{gorling}
and by Levy and Nagy \cite{levy1}. The theory is based on the 
constrained-search approach \cite{levy2} and proposes that the energy of an 
excited-state can also be written as a functional
\begin{equation}
E[\rho] = F[\rho,\rho_{0}] + \int\rho({\bf r})v_{ext}({\bf r})d{\bf r}
\label{1}
\end{equation}
of the excited-state density $\rho({\bf r})$. Here $F[\rho,\rho_{0}]$ is a
bi-density functional that depends on the ground-state density $\rho_{0}$
also, and $v_{ext}({\bf r})$ is the external potential that the electrons
are moving in. The bi-density functional for the density $\rho$ of the $nth$
excited-state is defined via the constrained-search formulation as
\begin{equation}
F[\rho,\rho_{0}] = {\rm min}_{\Psi \to \rho} \left< \Psi | {\hat T} + {\hat V}_{ee}|\Psi \right> \;,
\label{2}
\end{equation}
where $\Psi$ is orthogonal to the lower $(n-1)$ states of the hamiltonian, already
determined by the density $\rho_{0}$. Such a way of obtaining the functional
$F[\rho,\rho_{0}]$ makes it non-universal and also state-dependent. The
exchange-correlation energy functional $E_{xc}[\rho,\rho_{0}]$ for the
excited-state is then defined as the difference of $F[\rho,\rho_{0}]$ and
the non-interacting kinetic energy $T_{s}[\rho,\rho_{0}]$ corresponding to $\rho$. 
The latter is defined in a manner similar to Eq.~\ref{2} by dropping the operator
${\hat V}_{ee}$ from the right hand side.  Thus (for brevity, from here onwards
we drop $\rho_{0}$ from the argument of the functional)
\begin{equation}
E_{xc}[\rho] = F[\rho]-T_{s}[\rho]\;.
\label{3}
\end{equation}
With the assumption that the excited-state density is non-interacting v-representable,  
the density is obtained by solving the excited-states Kohn-Sham equation (atomic
units are used throughout the paper)
\begin{equation}
\left[-\frac{1}{2}\nabla^{2}+v_{ext}({\bf r}) + \int\frac{\rho({\bf r'})}
{|{\bf r}- {\bf r'}|}d{\bf r'}+v_{xc}({\bf r})\right]\phi_{i}({\bf r})=
\epsilon_{i}\phi_{i}
({\bf r})
\label{4}
\end{equation}
as
\begin{equation}
\rho({\bf r}) = \Sigma_{i} {\rm n}_{i}|\phi_{i}({\bf r})|^{2}\;,
\label{5}
\end{equation}
where ${\rm n}_{i}$ is the occupation number of the orbital $\phi_{i}$.
In Eq.~\ref{4} the various terms have their standard meaning with $v_{xc}({\bf r})$ 
representing the exchange-correlation potential for the excited-state. It is 
determined by taking the functional derivative of the excited-state 
exchange-correlation energy functional. That a Kohn-Sham like calculation can be 
performed for the excited-states was first proposed by Harbola and Sahni \cite{hs} 
on physical grounds, and has been put \cite{ssms} on a rigorous mathematical footing 
recently on the basis of differential virial theorem \cite{holas}. Calculations of 
excited-states energies based on the Harbola-Sahni work have yielded excellent results 
\cite{sen,deb}. The near exact exchange-correlation potential for the singlet 
$1s2s\;^{1}S$ excited-state of helium has also been constructed \cite{harb1} recently. 
However, we are not aware of any work where an exchange-correlation functional for the 
excited-states has been reported; In performing excited-state calculations 
\cite{gorling,levy1,harb2}, either the ground-state functionals or the orbital 
based-theories \cite{hs,opm} have been employed. The proposition for the construction
of an excited-state exchange-correlation functional is indeed a difficult one
since the functional is non-universal and also state-dependent.  Thus a general
functional form for it may not exist.

Against such a background, we ask if it is at all possible to obtain a simple
LDA-like functional for the excited-states.  To keep matters simple, we have
been looking at this problem within the exchange-only approximation. In this 
paper we show that it is indeed possible to construct an exchange
energy functional that gives transition energies comparable to the exact
exchange-only theories such as Hartree-Fock \cite{fischer}, optimized potential
\cite{opm} or the Harbola-Sahni \cite{sahnib} theory. The construction of the 
functional is based on the homogeneous electron-gas and in finding the final form 
of the functional we are guided mostly by qualitative plausibility arguments. Our 
work is thus exploratory in nature and represents probably the first attempt to 
construct an excited-state exchange-energy functional in terms of the density. 
The evidence of the accuracy of the functional constructed by us is given by the 
results of the transition energies of a large number of excited-states.  We also 
refer the reader to ref.~\cite{gorling2} for an expression for the change in the 
exchange energy in terms of the ground-state Kohn-Sham orbitals when an electron
is promoted from a lower energy orbital to a higher one.  

In the present work we take a particular class of excited states in which some
core orbital are filled, then there are some vacant orbitals and again there are 
some filled orbitals. We construct an LDA-like functional for such states in the
following section.

\section{Construction of the functional}

\begin{figure}[thb]
\includegraphics{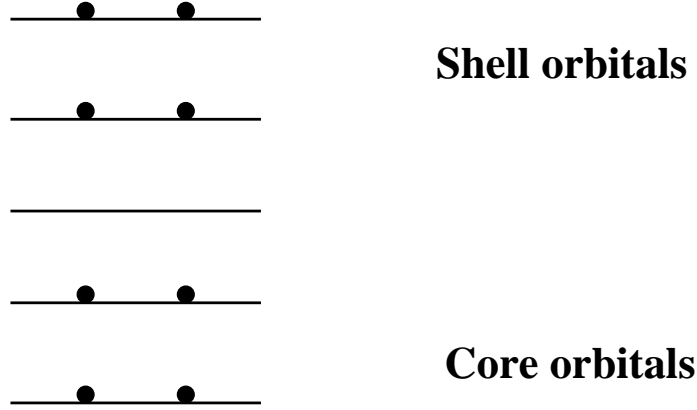}
\caption{Orbital occupation in an excited state configuration.}
\label{f:co-she}
\end{figure}

As stated above, we now consider such excited-states where the occupation of the 
orbitals is such that the electrons occupy some core orbitals and some shell orbitals, 
leaving the orbitals between the core and the shell region vacant.  This is shown 
schematically in Fig.~\ref{f:co-she}. 
Such an excited-state would be obtained, for 
example, if an electron from the filled orbitals of the ground-state is excited to just 
above the occupied levels. The exact exchange energy for a set of occupied orbitals 
is given as
\begin{equation}
E_{X} = -\frac{1}{2}\sum_{\sigma} \sum_{i}^{occ} \sum_j^{occ} \left<\phi_{i}
{({\bf r}_{1})}\phi_{j}{({\bf r}_{2})} \left|\frac{1}{ |{\bf r}_{1}-{\bf r}_{2}|} \right|\phi_{j}
{({\bf r}_{1})}\phi_{i}{({\bf r}_{2})} \right>
\label{6}
\end{equation}
so that the excited-state exchange energy when an electron is transferred
from one of the orbitals occupied in the ground-state to the lowest unoccupied
level is given as
\begin{eqnarray}
E_{X}^{excited} & = & E_{X}^{ground} + {\sum_{j}} \left<\phi_{rem}{({\bf r}_{1})}
     \phi_{j}{({\bf r}_{2}) } \left|\frac{1}{ |{\bf r}_{1}-{\bf r}_{2}|}\right|\phi_{j}
     {({\bf r}_{1})}\phi_{rem}{({\bf r}_{2})} \right> \nonumber \\
             &   & -\frac{1}{2}\int\int\frac{|\phi_{rem}({\bf r}_{1})|^{2}|\phi_{rem}
({\bf r}_{2})|^{2}}{|{\bf r}_{1}-{\bf r}_{2}|}d{\bf r}_{1}d{\bf r}_{2} 
         -\frac{1}{2}\int\int\frac{|\phi_{add}({\bf r}_{1})|^{2}|\phi_{add}
({\bf r}_{2})|^{2}}{|{\bf r}_{1}-{\bf r}_{2}|}d{\bf r}_{1}d{\bf r}_{2} \nonumber \\
                &   & - {\sum_{j(j\neq add)}} \left<\phi_{add}{({\bf r}_{1})}
     \phi_{j}{({\bf r}_{2}) } \left|\frac{1}{ |{\bf r}_{1}-{\bf r}_{2}|} \right|\phi_{j}
     {({\bf r}_{1})}\phi_{add}{({\bf r}_{2})} \right>\;, 
\label{7}
\end{eqnarray}
where $\phi_{rem}$ represents the orbital from which the electron has been removed 
and $\phi_{add}$ where the electron is added. The sum over the index $j$ in the 
second term on the right hand side runs over all the orbitals, including $\phi_{rem}$ 
and $\phi_{add}$, up to the highest occupied orbital in the excited-state.  On the other 
hand the sum in the fifth term runs over all the orbital except $\phi_{add}$.  We now 
attempt to make an LDA-like approximation for the excited-state exchange energy so that 
the difference (the last four terms in the equation above) between the approximate 
excited- and ground-state exchange energies is close to that given by the exact 
expression above. In making this approximation accurate, it is evident that the
self-energy terms (third and fourth terms on the right hand side of Eq.~\ref{7})
for the orbitals $\phi_{rem}$ and $\phi_{add}$ are to be treated accurately.

\begin{figure}[thb]
\includegraphics{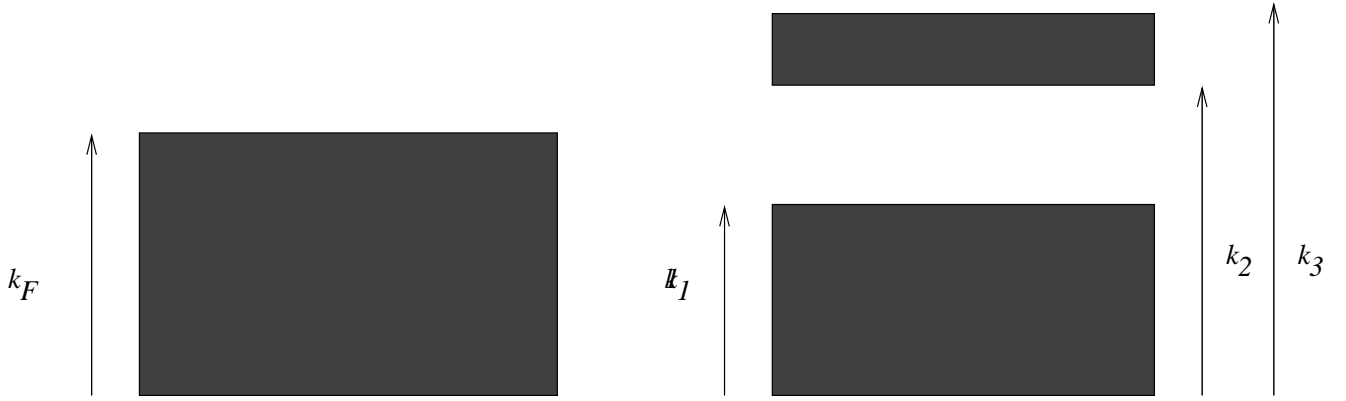}
\caption{$k-$space occupation in the ground and the excited state configuration.}
\label{k-space}
\end{figure}

As the first step towards an excited-state functional, we make the correspondence
between the excited-states that we are considering and similar excitations in a
homogeneous electron gas (HEG). If the HEG is in it's ground state, the electrons are  
filled up to the Fermi level so that the electrons occupy wave-vectors in k-space from 
$k=0$ to $k_{F}=(3\pi^{2}\rho)^{\frac{1}{3}}$, where $\rho$ is the electron density.
On the other hand, in an excited state of the system the electrons will occupy k-space 
differently compared to the ground state. For the kind of excited-states that we
consider in this paper, the corresponding occupation in the k-space is as follows:
The electrons occupy orbitals from $k=0$ to $k_{1}$ and $k_{2}$ to $k_{3}$ 
with a gap in between as shown in Fig.~(\ref{k-space}).
So that the excited state density is given by
\begin{equation}
k_{1}^{3} = 3\pi^{2}\rho_{c}\;,
\label{8}
\end{equation}
\begin{equation}
k_{3}^{3}-k_{2}^{3} = 3\pi^{2}\rho_{s}\;,
\label{9}
\end{equation}
and
\begin{equation}
k_{3}^{3}-k_{2}^{3}+k_{1}^{3} = 3\pi^{2}\rho
\label{10}
\end{equation}
where
\begin{equation}
\rho = \rho_{c}+\rho_{s}\;.
\label{11}
\end{equation}
In Eq.~\ref {8} $\rho_{c}$ and $\rho_{s}$ are the core (corresponding to the electrons 
occupying $k-$space from zero to $k_{1}$) and the shell (corresponding to the electrons 
occupying $k-$space from $k_{2}$ to $k_{3}$) density, respectively, and $\rho$ is
the total density.

The exchange energy for the HEG that occupies the k-space as described above
can be obtained exactly and is given as (MLDA stands for modified local density
approximation)
\begin{equation}
E_{X}^{MLDA}=E_{X}^{core}+E_{X}^{shell}+E_{X}^{core-shell}
\label{12}
\end{equation}
where
\begin{equation}
E_{X}^{core} = V\left[-\frac{k_{1}^{4}}{4\pi^{3}}\right]
\label{13}
\end{equation}
is the exchange energy of the core electrons,
\begin{equation}
E_{X}^{shell} = -\frac{V}{8\pi^{3}}\left[2(k_{3}^{3}-k_{2}^{3})(k_{3}-k_{2})
+ (k_{3}^{2}-k_{2}^{2})^{2}\;ln\left(\frac{k_{3}+k_{2}}{k_{3}-k_{2}}\right)\right]
\label{14}
\end{equation}
is the exchange energy of the electrons in the shell, and
\begin{eqnarray}
E_{X}^{core-shell} & = & - \frac{V}{8 \pi^3} \left[2 (k_3 - k_2) k_1^3+
2 (k_3^3 - k_2^3) k_1 + (k_2^2 - k_1^2)^2 \ln
\left(\frac{k_2 + k_1}{k_2 - k_1} \right) \right] \nonumber \\
 & & - \left(k_3^2 - k_1^2)^2 \ln \left(\frac{k_3 + k_1}{k_3 - k_1} \right) \right]
\label{15}
\end{eqnarray}
represents the exchange energy of interaction between the core and the shell
electrons. Here $V$ is the volume of the HEG.  After adding the three terms, the 
exchange-energy can also be written in the form
\begin{equation}
E_{X}^{MLDA} = \int\rho\left[\epsilon(k_{3}) - \epsilon(k_{2})
+\epsilon(k_{1})\right]d{\bf r} + log\; terms
\label{16}
\end{equation}
where $\epsilon(k)$ represents the exchange-energy per particle when the HEG is in its 
ground-state with the Fermi momentum equal to $k$. The equation above has a nice 
interpretation:  The integral on the right-hand side represents the exchange energy of 
the system of electrons with density $\rho$ when per electron energy is written as 
$[\epsilon(k_{3}) - \epsilon(k_{2}) +\epsilon(k_{1})]$, i.e. the per electron energy is 
given according to the occupation in the k-space (compare with Eq.~\ref{10}). The log 
terms, on the other hand, have no such simple interpretation. They have the kinetic 
energy density in them but we have not been able to write the terms in as easy a form as 
the first term. That the functional above has all the right limits if we take 
$k_{1}=k_{2}$ or $k_{2}=k_{3}$ is easily verified. Finally, the modified local spin 
density (LSD) functional $E_{X}^{MLSD}[\rho_{\alpha},\rho_{\beta}]$ in terms of the spin 
densities $\rho_{\alpha}$ and $\rho_{\beta}$ is easily obtained from the functional above 
as
\begin{equation}
E_{X}^{MLSD}[\rho_{\alpha},\rho_{\beta}] = \frac{1}{2}E_{X}^{MLDA}
[2\rho^{\alpha}] + \frac{1}{2}E_{X}^{MLDA}[2\rho^{\beta}] 
\label{lsda}
\end{equation}

Having derived the exchange functional for the HEG, we now apply it to the 
excited-states of various atoms to check if the functional above gives exchange
energy differences accurately. The excited-states chosen are such that they can be 
represented by a single Slater determinant so that the LDA is a good approximation 
\cite{ziegler,vonB} for them.  The different radii in the k-space, $k_{1}$, 
$k_{2}$ and $k_{3}$, needed to evaluate the exchange energy are found by Eqs.~
\ref{8}, \ref{9} and \ref{10}. For each state (ground and excited), the same set
of orbitals \cite{fnote1}  is employed to get the Hartree-Fock and the LSD exchange 
energies. We calculate the LSD and MLSD exchange energies using spherical spin 
densities since the effect of non-sphericity on the total exchange energy is small
\cite{janak}. 

In Table I we show the difference between the excited-state exchange energy and the
ground-state exchange energy for some atoms and ions. In the first column we give
the difference as obtained by the Hartree-Fock expression for the exchange energy.
In the second coulumn, the numbers are given for both the excited-state and the
ground-state exchange energies obtained by employing the ground-state LSD functional.
The third column gives the energy difference when the excited-state exchange energy
is calculated using the functional of Eq.~\ref{12}. It is clearly seen that the 
ground-state LSD approximation underestimates this energy difference.  This is
not surprising since the ground-state functional would give a larger exchange energy
for the excited-state than what a proper excited-state functional should give. However,
when the functional of Eq.~\ref{12} is employed to calculate the exchange energy for the
excited-states we find, to our surprise, that for the majority of the atoms the functional 
overestimates the differences by a large amount, whereas we expected to find the error to 
be about $10\%$ which is the general LDA exchange energy error. We note that this large
difference cannot come because we have spherical densities. If non-spherical densities
are used, the difference may increase even further.  For example, for the fluorine
atom, the ground-state exchange energy will become more negative for non-spherical
densities. On the other hand, the excited-state exchange energy will remain unchanged 
since the density is already spherical.  This will result in an even larger difference in
the exchange energies of the two states.
 
We now look for possible sources of error in the exchange-energy differences when
the functional of Eq.~\ref{12} is employed to get the exchange energy for the
excited-states. For this we examine Eq.~\ref{7} in which the last four terms on the
right hand side represent the exchange energy difference. Thus
\begin{eqnarray}
\Delta E_{X} & = &{\sum_{j}} \left<\phi_{rem}{({\bf r}_{1})}
     \phi_{j}{({\bf r}_{2}) } \left|\frac{1}{ |{\bf r}_{1}-{\bf r}_{2}|} \right|\phi_{j}
     {({\bf r}_{1})}\phi_{rem}{({\bf r}_{2})} \right> \nonumber \\
          &   & -\frac{1}{2}\int\int\frac{|\phi_{rem}({\bf r}_{1})|^{2}|\phi_{rem}
({\bf r}_{2})|^{2}}{|{\bf r}_{1}-{\bf r}_{2}|}d{\bf r}_{1}d{\bf r}_{2} 
                -\frac{1}{2}\int\int\frac{|\phi_{add}({\bf r}_{1})|^{2}|\phi_{add}
({\bf r}_{2})|^{2}}{|{\bf r}_{1}-{\bf r}_{2}|}d{\bf r}_{1}d{\bf r}_{2} \nonumber \\
                &   & - {\sum_{j(j\neq add)}} \left<\phi_{add}{({\bf r}_{1})}
     \phi_{j}{({\bf r}_{2}) } \left|\frac{1}{ |{\bf r}_{1}-{\bf r}_{2}|} \right|\phi_{j}
     {({\bf r}_{1})}\phi_{add}{({\bf r}_{2})} \right>\;.
\label{17}
\end{eqnarray}
It is the LDA values to this term that are given in Table I. The sources of error in 
this term we suspect are the self-exchange energies of the orbitals $\phi_{rem}$ and 
$\phi_{add}$ involved in the electron transfer. We now argue that the self-energy 
correction for both the terms sould be in the same direction. Thus for both the orbitals 
the self-interaction correction (SIC) is made by subtracting \cite{perdewz}
\begin{equation}
E_{X}^{SIC}[\phi] = \frac{1}{2}\int\int\frac{|\phi({\bf r}_{1})|^{2}
   |\phi({\bf r}_{2})|^{2})}{|{\bf r}_{1}-{\bf r}_{2}|}d{\bf r}_{1}d{\bf r}_{2}
   - E_{X}^{LSD}[\rho(\phi)]\;,
\label{18}
\end{equation}
where $\rho(\phi)$ is the orbital density for the orbital $\phi$. The argument goes
as follows.  The LDA should be reasonablty accurate when the integral over $k$ is
continuous. This makes the first term in the energy difference accurate.  We do 
have a choice of writing the first and the second terms as 
\begin{equation}
 {\sum_{j(j\neq rem)}} \left<\phi_{rem}{({\bf r}_{1})}
     \phi_{j}{({\bf r}_{2}) } \left|\frac{1}{ |{\bf r}_{1}-{\bf r}_{2}|} \right|\phi_{j}
     {({\bf r}_{1})}\phi_{rem}{({\bf r}_{2})} \right> 
                 +\frac{1}{2}\int\frac{|\phi_{rem}({\bf r}_{1})|^{2}|\phi_{rem}
({\bf r}_{2})|^{2}}{|{\bf r}_{1}-{\bf r}_{2}|}d{\bf r}_{1}d{\bf r}_{2} 
\label{19}
\end{equation}
but then the first term above will not be accurate as the LDA to it would involve
integration in k-space with a break from $k_{1}$ to $k_{2}$. Therefore to keep the
LDA accurate, we keep the summation continuous and write the self-interaction
energy of the electron removed with a negative sign in front.  By including the
self-interaction correction for the removed electron only, we find that the error
in the exchange energy difference reduces to about $10\%$ of the corresponding
HF value. To make the difference even more acuurate, we now consider the term for 
the orbital $\phi_{add}$ where the electron is added.  There the electron comes in 
with its self-interaction so for the added orbital too $E_{X}^{SIC}$ should be subtracted 
to make the results for the energy difference comparable to the Hartree-Fock results.
Thus the final expression for the exchange-energy that we have is
\begin{equation}
E_X^{MLSDSIC} = E_X^{MLSD}-E_{X}^{SIC}[\phi_{rem}]-E_{X}^{SIC}[\phi_{add}]
\label{20}
\end{equation}
where $E_X^{MLSD}$ is the energy functional given by Eq.~\ref{12} and 
$E_{X}^{SIC}[\phi]$ is given by Eq.~\ref{18}. We now compute the exchange energy 
differences given by the functional in Eq.~\ref{20} and show them in Table I. As 
is evident from the numbers displayed there, the functional of Eq.~\ref{20} gives 
highly accurate exchange-energy differences for all the systems considered.  When 
the exchnage-energy difference between the ground- and the excited-state is small, 
the HF, LSD and the functionals derived above, all give roughly the same results.  
However, when this difference is large, the LDA underestimates the magnitude of the 
difference by a large amount whereas the functional of Eq.~q:\ref{12} overestimates it.  
Only when the latter is corrected for the self-interaction is the difference almost 
the same as the Hartree-Fock difference.

Having obtained the functional to obtain accurate exchange energy difference, we now
apply it to a large number of excited-states of the class considered here and find
that we get the transition energies very close to those given by the Hartree-Fock 
theory.

\section{Results}
We now employ the exchange functional $E_{X}^{MLSDSIC}$ proposed above to obtain the 
transition energies for a variety of excitations in different atoms. We find that for 
all the systems the transition energies obtained by us are very close to the
corresponding Hartree-Fock energies \cite{fnote2}.  Our calculations proceed as 
follows: We get the ground-state energy by solving the Kohn-Sham equation with the 
effective exchange potential calculated using the Dirac formula \cite{dirac}.  We then 
solve the Kohn-Sham equation with the same (corresponding to the ground-state) functional 
for the excited-state configuration.  This gives us the excited-state energy $E_{LSD}$.
The difference between $E_{LSD}$ and the ground-state energy gives us the transition energy
$\Delta E_{LSD}$.  We then employ the Kohn-Sham orbitals from this calculation to get 
the modified LDA exchange energy including SIC by employing the functional $E_{X}^{MLSDSIC}$
of Eq.~\ref{19}. Making appropriate corrections we then get the total excited-state energy 
corresponding to this functional, and its difference with the ground-state energy gives 
us $\Delta E_{MLSDSIC}$. Although we have not performed self-consistent calculations with 
the new functional, self-consistency is not expected to affect the results significantly.
This is because, as we shall see in the results, the major difference in the transition
energies given by different functionals arises from the difference in the value of the
exchange energy itself.

Shown in Table II are the transition energies $\Delta E_{HF}$, $\Delta E_{LSD}$
and $\Delta E_{MLSDSIC}$ for some light atoms and ions when one of their inner 
electrons is excited to the lowest available orbital. The excitation energy
in these systems is such that for some of them $\Delta E_{LSD}$ is close to
$\Delta E_{HF}$ but for others it is not.  However, $\Delta E_{MLSDSIC}$ is
uniformly accurate for all the systems. We note that the error in $\Delta E_{LSD}$
is almost fully from the error in the corresponding exchange energy difference.
This is evident from a comparison of the numbers in Table I (for the
exchange energy differences) and in Table II.  

In Table III we look at the excitation energies of the alkali atoms and $Mg^{+}$
by exciting an electron from the uppermost orbital to an outer orbital.  These
are weakly bound systems and as such their excitation energies are relatively
smaller.  Thus they provide a good testing ground for the proposed functional.
An interesting point about these systems is that the LSD itself gives excitation
energies close to the HF excitation energies. It is therefore quite gratifying to
see that the transition energies obtained by the new functional also 
are of very good quality, although the present method tends to slightly overestimate the
transition energies.

Next we consider some bigger atoms where we can excite the electron from
more than one inner orbital.  Shown in Tables IV and V are the excitation energies
for the atoms in the third row of the periodic table. In Table IV we consider
an electron being excited from the $3s$ orbital to the $3p$ orbital. In all
these case $\Delta E_{LSD}$ is smaller than the true energy difference
whereas the present functional gives highly accurate estimates of the transition
energy. Notice again that the error in the value of $\Delta E_{LSD}$ arises 
mainly from the error in the exchange energy.

In Table V, we show the transition energies for the same set of atoms and ions
as in Table IV, but for the electron now being excited from the $2s$ orbital
to the $3p$ orbital.  Consequently the energy of excitation is much larger
in this case. The LSD in all these cases underestimates the excitation energy,
whereas the present functional gives accurate results although slightly
overestimating them.  However, the error with respect to the LSD is reduced by
a factor of $5$ or more.

Shown in Table VI are the excitation energies for a group of atoms for which
the LSD gives transition energies very close to the HF excitation energies.
In all the cases we find that the functional proposed here is able to give
accurate excitation energies.  Thus we find that when the LSD results are
accurate, so are the results given by the new functional.  What is significant,
however, is that when the LSD results are poor, the new functional properly
corrects the error in LSD.

Finally, we consider the cases where two electrons are excited to the higher orbitals. 
As already pointed out, double excitations are difficult to deal with in the TDDFT approach 
to finding excitation energies. Results for different excitations for a variety of atomic 
systems are shown in Table VII.  As is evident from the table, for all the systems, our 
method gives excellent results whereas the LSD underestimates the energies. 

\section{Discussion and concluding remarks}
In the above we have presented a new LDA-like functional for obtaining the excitation
energies.  It has been employed to investigate over $40$ excited states. The results
show that our procedure gives accurate excitation energies for all of them, whereas for 
most of the systems the LSD underestimates the energy difference.  We have worked within 
the exchange-only approximation and have chosen a particular class of excited-states.  
What we have learnt through the study reported here is that a simple extension of the 
LDA to the excited-states overestimates the energy differences.  This is due to the 
self-interaction of the electron inherent in the LDA.  When corrected for the 
self-interaction through a careful analysis, the resulting functional gives highly 
accurate answers for the excited-states.  Thus if more accurate functionals than the 
LDA are employed, our method of developing excited-state functionals should give good 
excited-state functionals. 

In the present work, we have developed a functional for a particular class of
excited states and demonstrated that it is possible to construct excited-state
energy functionals that are capable of giving transition energies close to the 
exact theory. We are now working on functionals for states other than considered
in this paper.  As pointed out in the introduction, excited-state functionals are
not universal and therefore have to be dealt with separately for different kinds
of excited-states.

In this work, we have also not looked at the correlation energy functionals.  Can
correlation energy functionals be developed along similar lines?  We trust that
it should be possible and are working on this problem.

Finally, we also wish to look at the ultimate excitation i.e. the ionization of 
the system. If the electron is removed from the uppermost orbital, the LDA is known to 
give good ionization energies when calculated through the total energy difference.
In these cases too our functional would give results the same as those obtained
from the LDA: If we go through the arguments presented earlier, we find that in
these cases our functional reduces to the LDA functional for the core electrons. This
is because in calculating the exchnage energy, the summation over the occupied
orbitals is continuous and therefore we do not have to make any self-interaction 
correction for the removed electron.  Further the shell electron density vanishes
so the contribution from the added electron is zero.  Interestingly, we find that
if we ionize the atom by removing an electron from the inner orbitals, we obtain
accurate removal energies if we think of the process in two spteps - first removing
an electron from the top level and then exciting the resulting ion by exciting
an electron from the inner orbital to the top of filled orbitals. More work 
on such different kinds of excitation is in progress and will be reported in the
future.

{\bf Acknowledgement:} We thank Professor K.D. Sen for providing the Hartree-Fock
data on excited-states of atoms. Comments of Rajan Pandey on the manuscript are
appreciated.

\newpage
\begin{table}
\caption{Difference in the exchange energies of the ground- and excited-states of 
some atoms and ions.  The First column gives the atom/ion and the transition, the 
second column the difference $\Delta E_X^{HF}$ as obtained in Hartree-Fock theory, the 
third column the difference $\Delta E_X^{LSD}$ given by the ground-state energy 
functional.  The fourth and the fifth column describe the difference as obtained with the 
functional proposed in this paper.  The fourth column gives the exchange-energy difference 
$\Delta E_X^{MLSD}$ obtained by employing the functional of Eq.~\ref{12} whereas the fifth
column gives that given by the functional of Eq.~\ref{19} $\Delta E_X^{MLSDSIC}$. Numbers
given are in atomic units.}
\vspace{0.2in}
\begin{tabular}{lcccc}
\hline
atoms/ions&$\Delta$$E_X^{HF}$&$\Delta$$E_X^{LSD}$&$\Delta$$E_X^{MLSD}$&
$\Delta$$E_X^{MLSDSIC}$\\
\hline
$Li(2s^{1}\;^{2}S\rightarrow2p^{1}\;^{2}P)$ &0.0278&0.0264&0.0587&0.0282 \\
$B(2s^{2}2p^{1}\;^{2}P\rightarrow2s^{1}2p^{2}\;^{2}D)$ &0.0353&0.0319&0.0998&0.0412 \\
$C(2s^{2}2p^{2}\;^{3}P\rightarrow2s^{1}2p^{3}\;^{3}D)$ &0.0372&0.0332&0.1188&0.0454 \\
$N(2s^{2}2p^{3}\;^{4}S\rightarrow2s^{1}2p^{4}\;^{4}P)$ &0.0399&0.0353&0.1381&0.0503 \\
$O(2s^{2}2p^{4}\;^{3}P\rightarrow2s^{1}2p^{5}\;^{3}P)$ &0.1582&0.0585&0.2634&0.1624 \\
$F(2s^{2}2p^{5}\;^{2}P\rightarrow2s^{1}2p^{6}\;^{2}S)$ &0.3021&0.0891&0.3908&0.2765 \\
$Ne^{+}(2s^{2}2p^{5}\;^{2}P\rightarrow2s^{1}2p^{6}\;^{2}S)$ &0.3339&0.0722&0.4397&0.3037 \\
$S(3s^{2}3p^{4}\;^{3}P\rightarrow3s^{1}3p^{5}\;^{3}P)$ &0.1106&0.0475&0.1798&0.1252 \\
$Cl^{+}(3s^{2}3p^{4}\;^{3}P\rightarrow3s^{1}3p^{5}\;^{3}P)$ &0.1257&0.0483&0.2050&0.1441\\
$Cl(3s^{2}3p^{5}\;^{2}P\rightarrow3s^{1}3p^{6}\;^{2}S)$ &0.2010&0.0603&0.2567&0.1969\\
\hline
\end{tabular}
\end{table}

\begin{table}
\caption{Transition energies, in atomic units, of an electron being excited from the 
$2s$ orbital of some atoms to their $2p$ orbital.  The first column gives this energy
as obtained in Hartree-Fock theory.  The numbers in the second column are obtained by 
employing the ground-state LDA for both the ground- and the excited-state.  The last 
column gives the energies given by employing the ground-state LDA for the ground-state 
and the functional of Eq.~\ref{19} for the excited-state.}
\vspace{0.2in}
\begin{tabular}{lccc}
\hline
atoms/ions & $\Delta$$E_{HF}$ & $\Delta$E(LSD) &   $\Delta$E(MLSDSIC)  \\
\hline
$N(2s^{2}2p^{3}\;^{4}S\rightarrow2s^{1}2p^{4}\;^{4}P)$ &0.4127&0.3905&0.4014 \\
$O^{+}(2s^{2}2p^{3}\;^{4}S\rightarrow2s^{1}2p^{4}\;^{4}P)$ &0.5530&0.5397&0.5571 \\
$O(2s^{2}2p^{4}\;^{3}P\rightarrow2s^{1}2p^{5}\;^{3}P)$ &0.6255&0.5243&0.6214 \\
$F^{+}(2s^{2}2p^{4}\;^{3}P\rightarrow2s^{1}2p^{5}\;^{3}P)$ &0.7988&0.6789&0.8005 \\
$F(2s^{2}2p^{5}\;^{2}P\rightarrow2s^{1}2p^{6}\;^{2}S)$ &0.8781&0.6671&0.8573 \\
$Ne^{+}(2s^{2}2p^{5}\;^{2}P\rightarrow2s^{1}2p^{6}\;^{2}S)$ &1.0830&0.8334&1.0607 \\
\hline
\end{tabular}
\end{table}

\begin{table}
\caption{The caption is the same as that for Table I except that we are now
considering transitions from the outermost orbital to an upper orbital for
weakly bound systems.}
\vspace{0.2in}
\begin{tabular}{lccc}
\hline
atoms/ions & $\Delta$$E_{HF}$ & $\Delta$E(LSD) &   $\Delta$E(MLSDSIC)  \\
\hline
$Li(2s^{1}\;^{2}S\rightarrow2p^{1}\;^{2}P)$ &0.0677&0.0646&0.0672 \\
$Na(3s^{1}\;^{2}S\rightarrow3p^{1}\;^{2}P)$ &0.0725&0.0751&0.0753 \\
$Mg^{+}(3s^{1}\;^{2}S\rightarrow3p^{1}\;^{2}P)$ &0.1578&0.1585&0.1696 \\
$K(4s^{1}\;^{2}S\rightarrow4p^{1}\;^{2}P)$ &0.0516&0.0556&0.0580 \\
\hline
\end{tabular}
\end{table}

\begin{table}
\caption{Electron transition energy from the $3s$ to the $3p$ orbital in some atoms.}
\vspace{0.2in}
\begin{tabular}{lccc}
\hline
atoms/ions & $\Delta$$E_{HF}$ & $\Delta$E(LSD) &   $\Delta$E(MLSDSIC)  \\
\hline
$P(3s^{2}3p^{3}\;^{4}S\rightarrow3s^{1}3p^{4}\;^{4}P)$ &0.3023&0.2934&0.3055 \\
$S(3s^{2}3p^{4}\;^{3}P\rightarrow3s^{1}3p^{5}\;^{3}P)$ &0.4264&0.3615&0.4334 \\
$Cl^{+}(3s^{2}3p^{4}\;^{3}P\rightarrow3s^{1}3p^{5}\;^{3}P)$ &0.5264&0.4482&0.5403 \\
$Cl(3s^{2}3p^{5}\;^{2}P\rightarrow3s^{1}3p^{6}\;^{2}S)$ &0.5653&0.4301&0.5630 \\
$Ar^{+}(3s^{2}3p^{5}\;^{2}P\rightarrow3s^{1}3p^{6}\;^{2}S)$ &0.6769&0.5174&0.6766 \\
\hline
\end{tabular}
\end{table}

\begin{table}
\caption{Electron transition energy from the $2s$ to the $3p$ orbital in the
same atoms as in Table IV.}
\vspace{0.2in}
\begin{tabular}{lccc}
\hline
atoms/ions & $\Delta$$E_{HF}$ & $\Delta$E(LSD) &   $\Delta$E(MLSDSIC)  \\
\hline
$P(2s^{2}3p^{3}\;^{4}S\rightarrow2s^{1}3p^{4}\;^{4}P)$ &6.8820&6.4188&6.9564 \\
$S(2s^{2}3p^{4}\;^{3}P\rightarrow2s^{1}3p^{5}\;^{3}P)$ &8.2456&7.7337&8.3271 \\
$Cl^{+}(2s^{2}3p^{4}\;^{3}P\rightarrow2s^{1}3p^{5}\;^{3}P)$ &9.8117&9.2551&9.8997 \\
$Cl(2s^{2}3p^{5}\;^{2}P\rightarrow2s^{1}3p^{6}\;^{2}S)$ &9.7143&9.1653&9.8171 \\
$Ar^{+}(2s^{2}3p^{5}\;^{2}P\rightarrow2s^{1}3p^{6}\;^{2}S)$ &11.3926&10.8009&11.5061 \\
\hline
\end{tabular}
\end{table}

\begin{table}
\caption{Electron transition energies for atoms where LSD transition energies 
are accurate}
\vspace{0.2in}
\begin{tabular}{lccc}
\hline
atoms/ions & $\Delta$$E_{HF}$ & $\Delta$E(LSD) &   $\Delta$E(MLSDSIC)  \\
\hline
$B(2s^{2}2p^{1}\;^{2}P\rightarrow2s^{1}2p^{2}\;^{2}D)$ &0.2172&0.1993&0.2061 \\
$C^{+}(2s^{2}2p^{1}\;^{2}P\rightarrow2s^{1}2p^{2}\;^{2}D)$ &0.3290&0.3078&0.3216 \\
$C(2s^{2}2p^{2}\;^{3}P\rightarrow2s^{1}2p^{3}\;^{3}D)$ &0.2942&0.2878&0.2967 \\
$N^{+}(2s^{2}2p^{2}\;^{3}P\rightarrow2s^{1}2p^{3}\;^{3}D)$ &0.4140&0.4149&0.4305 \\
$Si^{+}(3s^{2}3p^{1}\;^{2}P\rightarrow3s^{1}3p^{2}\;^{2}D)$ &0.2743&0.2632&0.2799 \\
$Si(3s^{2}3p^{2}\;^{3}P\rightarrow3s^{1}3p^{3}\;^{3}D)$ &0.2343&0.2356&0.2442 \\
\hline
\end{tabular}
\end{table}

\begin{table}
\caption{Excitation energies of some atoms when two electrons are excited.}
\vspace{0.2in}
\begin{tabular}{lccc}
\hline
atoms/ions & $\Delta$$E_{HF}$ & $\Delta$E(LSD) &   $\Delta$E(MLSDSIC)  \\
\hline
$Be(2s^{2}\;^{1}S\rightarrow2p^{2}\;^{1}D)$ &0.2718&0.2538&0.2655 \\
$B(2s^{2}2p^{1}\;^{2}P\rightarrow2p^{3}\;^{2}D)$ &0.4698&0.4117&0.4798 \\
$C^{+}(2s^{2}2p^{1}\;^{2}P\rightarrow2p^{3}\;^{2}D)$ &0.6966&0.6211&0.7180 \\
$C(2s^{2}2p^{2}\;^{3}P\rightarrow2p^{4}\;^{3}P)$ &0.7427&0.5950&0.7312 \\
$N^{+}(2s^{2}2p^{2}\;^{3}P\rightarrow2p^{4}\;^{3}P)$ &1.0234&0.8369&1.0143 \\
$N(2s^{2}2p^{3}\;^{4}S\rightarrow2p^{5}\;^{2}P)$ &1.1789&0.9440&1.1785 \\
$O^{+}(2s^{2}2p^{3}\;^{4}S\rightarrow2p^{5}\;^{2}P)$ &1.5444&1.2552&1.5480 \\
$O(2s^{2}2p^{4}\;^{3}P\rightarrow2p^{6}\;^{1}S)$ &1.5032&1.1333&1.4736 \\
$F^{+}(2s^{2}2p^{4}\;^{3}P\rightarrow2p^{6}\;^{1}S)$ &1.8983&1.4381&1.8494 \\
$Mg(3s^{2}\;^{1}S\rightarrow3p^{2}\;^{1}D)$ &0.2578&0.2555&0.2651 \\
$S(3s^{2}3p^{4}\;^{3}P\rightarrow3p^{6}\;^{1}S)$ &1.0273&0.7807&1.0266 \\
$P(3s^{2}3p^{3}\;^{4}S\rightarrow3p^{5}\;^{2}P)$ &0.8539&0.6927&0.8680 \\
$Si^{+}(3s^{2}3p^{1}\;^{2}P\rightarrow3p^{3}\;^{2}D)$ &0.5856&0.5377&0.6230 \\
$Si(3s^{2}3p^{2}\;^{3}P\rightarrow3p^{4}\;^{3}P)$ &0.5860&0.4928&0.5986 \\
$Cl^{+}(3s^{2}3p^{2}\;^{3}P\rightarrow3p^{4}\;^{3}P)$ &1.2535&0.9551&1.2516 \\
\hline
\end{tabular}
\end{table}


\begin{thebibliography}{unsrt}

\bibitem{gross1} R.M. Dreizler and E.K.U. Gross, {\it Density Functional
Theory} , (Springer-Verlag, Berlin, 1990)
\bibitem{parr1} R.G. Parr and W. Yang, {\it Density-Functional Theory of
Atoms and Molecules}, (Oxford University Press, Oxford, 1999).
\bibitem{ziegler}T. Ziegler, A.Rauk and E.J. Baerends, Theor. Chim. Acta
{\bf 43}, 261 (1977).
\bibitem{Gunnar} O. Gunnarsson and B.I. Lundquist, Phys. Rev. B {\bf 13},
4274 (1976).
\bibitem{vonB} U. von Barth, Phys. Rev. A {\bf 20}, 1693 (1979).
\bibitem{LP1} J.P. Perdew and M. Levy, Phys. Rev. B {\bf 31}, 6264 (1985).
\bibitem{rkp} R.K. Pathak, Phys. Rev. A {\bf 29}, 978 (1984).
\bibitem{theo} A.K. Theophilou, J. Phys. C {\bf 12}, 5419 (1979).
\bibitem{olivi1} E.K.U. Gross, L.N. Oliviera and W. Kohn, Phys. Rev. A 
{\bf 37}, 2809 (1988); L.N. Oliviera, E.K.U. Gross and W. Kohn, Phys. Rev. A
{\bf 37}, 2821 (1988).
\bibitem{nagy} \'{A}. Nagy, J. Phys. B {\bf 29}, 389 (1996).
\bibitem{gross2} E. Runge and E.K.U. Gross, Phys. Rev. Lett. {\bf 52}, 997
(1984).
\bibitem{casida} M.E. Casida, {\it Recent Advances in Density Functional 
Methods, Part 1}, edited by D.P. Chong (World Scientific, Singapore,1995).
\bibitem{peters} M. Petersilka, U.J. Gossmann and E.K.U Gross, Phys. Rev. 
Lett. {\bf 76}, 1212(1996).
\bibitem{burke} N.T. Maitra, F. Zhang, R.J. Cave and K. Burke, J. Chem. Phys.
{\bf 120}, 5932 (2004).
\bibitem{gorling} A. G\"{o}rling, Phys. Rev. A {\bf 59}, 3359 (1999).
\bibitem{levy1} M. Levy and \'{A}. Nagy, Phys. Rev. Lett. {\bf 83}, 4361
(1999); \'{A}. Nagy and M. Levy, Phys. Rev. A {\bf 63}, 052502 (2001). 
\bibitem{levy2} M. Levy, Proc. Natl. Acad. Sci. USA {\bf 76}, 6062 (1979).
\bibitem{hs} M.K. Harbola and V. Sahni, Phys. Rev. Lett. {\bf 62}, 489
(1989).
\bibitem{ssms} V. Sahni, L. Massa, R. Singh  and M. Slamet, Phys. Rev. Lett.
{\bf 87}, 113002 (2001).
\bibitem{holas} A. Holas and N.H. March, Phys. Rev. A {\bf 55}, 2040 (1995).
\bibitem{sen} K.D. Sen, Chem. Phys. Lett. {\bf 188}, 510 (1992).
\bibitem{deb} R. Singh and B.M. Deb, Phys. Rep. {\bf 311}, 47 (1999).
\bibitem{harb1} M.K. Harbola, Phys. Rev. A {\bf 69}, 042512 (2004).
\bibitem{harb2} M.K. Harbola, Phys. Rev. A {\bf 65}, 052504 (2002).
\bibitem{opm} R.T. Sharp and G.K. Horton, Phys. Rev. {\bf 90}, 3876
(1953); J.D. Talman and W.F. Shadwick, Phys. Rev. A {\bf 14}, 36 (1976).
\bibitem{fischer} C.F. Fischer, {\it The Hartree-Fock method for atoms},
(John Wiley, New York 1977).
\bibitem{sahnib} V. Sahni, {\it Quantal density functional theory} 
(Springer Berlin, 2004).
\bibitem{gorling2} A. G\"{o}rling, Phys. Rev. A {\bf 54}, 3912 (1996).
\bibitem{fnote1} The orbitals employed are those obtained by solving
the Kohn-Sham equations with the Harbola-Sahni exchange potential (ref.
\cite{hs} above). These orbitals are very close (see, for example,
V. Sahni, Y. Li and M.K. Harbola, Phys. Rev. A {\bf 45}, 1434 (1992) and 
refs. \cite{sen,deb} above) to the Hartree-Fock orbitals.
\bibitem{janak} J.F. Janak and A.R. Williams, Phys. Rev. B {\bf 23}, 6301 (1981).
\bibitem{perdewz} J.P. Perdew and A. Zunger, Phys. Rev. B {\bf 23}, 5048 (1981).
\bibitem{fnote2} The energies are calculated by solving the Kohn-Sham equation
with the Harbola-Sahni potential. The resulting multiplet energies are essentially
the same (see ref. \cite{deb}) as those of Hartree-Fock theory. 
\bibitem{dirac} P.A.M. Dirac, Proc. Cambridge Phil. Soc. {\bf 26} ,
376 (1930).

\end{thebibliography}
\end{document}